\documentclass[twocolumn]{aastex631}

\usepackage{amsmath,amsfonts,amssymb}
\usepackage{fancyhdr}
\usepackage{graphicx}
\usepackage{xspace}
\usepackage{rotating}
\usepackage[normalem]{ulem}
\usepackage{braket}
\usepackage{verbatim}
\usepackage{xcolor}
\usepackage{hyperref}
\usepackage[utf8]{inputenc}

\def\lsim{\mathrel{\raise.3ex\hbox{$<$\kern-.75em\lower1ex\hbox{$\sim$}}}}
\def\gsim{\mathrel{\raise.3ex\hbox{$>$\kern-.75em\lower1ex\hbox{$\sim$}}}}

\newcommand{\be}{\begin{equation}}
\newcommand{\ee}{\end{equation}}
\newcommand{\bea}{\begin{equation}\begin{aligned}}
\newcommand{\eea}{\end{aligned}\end{equation}}

\usepackage{hyperref}
\usepackage{wrapfig}

\def\pfrac#1#2{\left( \frac{#1}{#2} \right)}
\def\avg#1{\langle #1 \rangle}

\def\iso#1#2{\mbox{${}^{#2}{\rm #1}$}}
\def\be1#1{\iso{Be}{1#1}}
\def\al2#1{\iso{Al}{2#1}}
\def\ar3#1{\iso{Ar}{3#1}}
\def\ca4#1{\iso{Ca}{4#1}}
\def\k4#1{\iso{K}{4#1}}
\def\mn5#1{\iso{Mn}{5#1}}
\def\fe6#1{\iso{Fe}{6#1}}
\def\rb8#1{\iso{Rb}{8#1}}
\def\nb9#1{\iso{Nb}{9#1}}
\def\zr9#1{\iso{Zr}{9#1}}
\def\tc9#1{\iso{Tc}{9#1}}
\def\pd10#1{\iso{Pd}{10#1}}
\def\i12#1{\iso{I}{12#1}}
\def\cs13#1{\iso{Cs}{13#1}}
\def\sm14#1{\iso{Sm}{14#1}}
\def\gd15#1{\iso{Gd}{15#1}}
\def\dy15#1{\iso{Dy}{15#1}}
\def\hf18#1{\iso{Hf}{18#1}}
\def\gd15#1{\iso{Gd}{15#1}}
\def\pb20#1{\iso{Pb}{20#1}}
\def\bi21#1{\iso{Bi}{21#1}}
\def\u23#1{\iso{U}{23#1}}
\def\np23#1{\iso{Np}{23#1}}
\def\pu24#1{\iso{Pu}{24#1}}
\def\cm24#1{\iso{Cm}{24#1}}
\def\th23#1{\iso{Th}{23#1}}
\def\re18#1{\iso{Re}{18#1}}

\def\apjl{ApJL}

\def\apjs{ApJS}

\def\aap{A\&A}

\begin{document}

\title{Proposed Lunar Measurements of $r$-Process Radioisotopes to Distinguish the Origin of Deep-sea \pu244}

\correspondingauthor{Xilu Wang}
\email{wangxl@ihep.ac.cn, xlwang811@gmail.com} 

\author[0000-0002-5901-9879]{Xilu Wang}
\affil{Key Laboratory of Particle Astrophysics, Institute of High Energy Physics, Chinese Academy of Sciences,  Beijing, 100049, China}
\affil{Department of Physics, University of California, Berkeley, CA 94720, USA}
\affil{Department of Physics, University of Notre Dame, Notre Dame, IN 46556, USA}
\affil{N3AS Collaboration}

\author[0000-0002-2881-7982]{Adam M. Clark}
\affiliation{Department of Physics, University of Notre Dame, Notre Dame, IN 46556, USA}

\author[0000-0002-7399-0813]{John Ellis}
\affiliation{Theoretical Physics and Cosmology Group, Department of Physics, King's College London, London WC2R 2LS, UK}
\affiliation{NICPB, R\"avala pst.~10, 10143 Tallinn, Estonia; Theoretical Physics Department, CERN, CH-1211 Geneva 23, Switzerland}

\author[0000-0002-3876-2057]{Adrienne F. Ertel}
\affiliation{Department of Astronomy, University of Illinois, Urbana, IL 61801, USA}
\affiliation{Illinois Center for the Advanced Study of the Universe, University of Illinois, Urbana, IL 61820}

\author[0000-0002-4188-7141]{Brian D. Fields}
\affiliation{Department of Astronomy, University of Illinois, Urbana, IL 61801, USA}
\affiliation{Department of Physics, University of Illinois, Urbana, IL 61801, USA}
\affiliation{Illinois Center for the Advanced Study of the Universe, University of Illinois, Urbana, IL 61820}

\author[0000-0002-2786-5667]{Brian J. Fry}
\affiliation{Department of Physics, United States Air Force Academy, Colorado Springs, CO 80840, USA}

\author[0000-0002-8056-2526]{Zhenghai Liu}
\affiliation{Department of Astronomy, University of Illinois, Urbana, IL 61801, USA}
\affiliation{Illinois Center for the Advanced Study of the Universe, University of Illinois, Urbana, IL 61820}

\author[0000-0001-5071-0412]{Jesse A. Miller}
\affiliation{Department of Astronomy, University of Illinois, Urbana, IL 61801, USA}
\affiliation{Illinois Center for the Advanced Study of the Universe, University of Illinois, Urbana, IL 61820}
\affiliation{Center for Space Physics, Department of Astronomy, Boston University, Boston, MA 02215, USA}

\author[0000-0002-4729-8823]{Rebecca Surman}
\affiliation{Department of Physics, University of Notre Dame, Notre Dame, IN 46556, USA}
\affiliation{N3AS Collaboration}

\begin{abstract}
\pu244 has recently been discovered in deep-sea deposits spanning the past 10 Myr, a period that includes two \fe60 pulses from nearby supernovae. \pu244 is among the heaviest $r$-process products, and we consider whether it 
was created in the supernovae, which is disfavored by nucleosynthesis simulations, or in an earlier kilonova event that seeded the nearby interstellar medium with \pu244 that was subsequently swept up by the supernova debris.  We discuss how these possibilities can be probed by measuring \pu244 and other {\em r}-process radioisotopes such as \i129 and \hf182, both in lunar regolith samples returned to Earth by missions such as {\it Chang'e} and {\it Artemis}, and in deep-sea deposits.
\end{abstract}

\keywords{ $r$-Process ; Supernovae ; Compact binary stars ; Lunar regolith ; Mass Spectrometry} 

\section{Introduction} 
\label{sec:intro}

Measurements of live radioactive isotopes can provide insights into recent astrophysical
explosions such as core-collapse supernovae (SNe) within ${\cal O}(100)$~pc of Earth~\citep{Ellis1996}
that are expected to occur every few million years, clarifying 
the possibility of rarer events within ${\cal O}(10)$~pc that might have caused
mass extinctions in the past~\citep{Ruderman1974,Ellis1995}. Many experiments over the past two decades
have detected pulses of live \fe60 in deep-sea deposits~\citep{Knie1999, Knie2004, Fitoussi2008, Wallner2016, Ludwig2016,Wallner2020}
from between 2 and 3 Myr ago (Mya), very likely due to a nearby core-collapse SN. 
There have also been measurements of \fe60 in the lunar regolith~\citep{Fimiani2016}, in cosmic rays~\citep{Binns2016}, 
and in Antarctic snow~\citep{Koll2019}. 

These measurements were accompanied by
some tantalizing hints of deep-sea \pu244~\citep{Paul2001,Wallner2004,Raisbeck2007,Wallner2015}.
These are interesting because the \pu244 
is produced exclusively 
by the astrophysical rapid neutron-capture process ($r$-process),
which is one of the dominant avenues to synthesize elements heavier than iron in the universe \citep{B2FH}.
The nature of the $r$-process lies at the heart of multimessenger astronomy, with connections to gravitational-wave observations and gamma-ray bursts \citep{GW170817, Abbott2017, Cowperthwaite2017, Kasen2017}, as well as observations of the most ancient stars \citep{Ian,Erika,Placco2020}.  The nuclear physics of the $r$-process demands that an intense neutron flux act over a short timescale $\sim 1 \ \rm s$; this points to explosive conditions \citep{B2FH}. The astrophysical sites that most plausibly host such conditions are core-collapse SNe and neutron-star mergers (kilonovae (KNe)): see the reviews by~\cite{Cowan1991, solar, Cowan2021, Kajino2019}, and references therein.  Which of these dominated production in the early Galaxy, and which dominates today, remains frustratingly difficult to identify unambiguously, in part because 
$r$-process observables such as abundance patterns potentially sum contributions from multiple events.

Deep-sea samples open a new window into the $r$-process and are particularly exciting because they give information on specific isotopes (not elemental sums),
and they sample ejecta from specific events \citep{Wang2021}. Comparing the early solar system abundance of \pu244 with the
deep-sea sediments from \cite{Wallner2015}, \cite{Hotokezaka2015} inferred that the measured \pu244 came from rare $r$-process events.
Motivated by these hints, in an earlier paper we studied possible signatures of rare SN and KN $r$-process
events, analyzing the potential implications of \pu244 detection, estimating the strengths of other
$r$-process radioisotope signatures, and discussing how they could help distinguish between potential sites~\citep{Wang2021}. 

A major advance in studies of live astrophysical radioisotopes has recently been made by~\cite{Wallner2021}
with the discovery 
\pu244 in deep-sea
ferromanganese (Fe-Mn) crusts from periods that include both this and the $\sim 3$~Mya \fe60 pulse. 
Whereas earlier hints had reported single \pu244 atoms, \cite{Wallner2021} reported $181\pm 19$ atoms above background, indicating a firm detection.
This same study also found an earlier \fe60 pulse $\sim 7$~Mya. 
These results add new dimensions to our picture of recent near-Earth explosions
and widen the scope of their implications.
Broadly, the detection of \pu244 is not only the second firmly detected radioisotope
in this epoch but it also demands an {\em r}-process source and so probes the astrophysical
site of the {\em r}-process.
Further, the second \fe60 pulse shows that there were multiple explosions,
as one would expect from massive stars that are  highly clustered \citep{Zinnecker2007}.
We thus follow \cite{Ertel2022} in referring to the event around 3 Mya as the Pliocene event (SN Plio) and the event around 7 Mya as the Miocene event (SN Mio).
In this paper, we
study the interpretation and potential implications of these new experimental results,
focusing on the information to be gained from lunar measurements of $r$-process radioisotopes.

As has been shown in \cite{Fry2015}, 
ordinary (non-{\em r}-process) nucleosynthesis  
in core-collapse SNe
provides the only plausible source of \fe60 observed in 
the two pulses.  
\cite{Wallner2021} concurs, making this a starting point for their analysis.
The question then becomes:
{\it Could either or both of these SNe also have produced the \pu244 or is a separate event required, presumably a KN?}

The two \fe60 pulses require at least two distinct SNe.
The \pu244 data
were not sampled as finely in time as the \fe60 data but in three broad time windows
including a surface layer that 
includes anthropogenic contamination.
The two deeper layers each overlap with a \fe60 pulse,
with similar \fe60/\pu244 ratios, represented by the yellow band in Fig.~\ref{fig:FePu}. The data show \fe60 to be much more abundant than \pu244
in both pulses.

Motivated by these data, \citet{Wang2021} proposed two scenarios for \pu244 deposition on the Earth by recent nearby events: (1) {\em one step}, i.e., the deposition of \pu244 is a direct consequence of the propagation of the ejecta from the nearby explosion events; and (2) {\em two-step} process, in which an earlier KN ejecta propagates to and mixes into the proto-Local Bubble, followed by the relative motion of the Earth and $r$-process-enriched dust, leading to the subsequent bombardment of \pu244 onto Earth.  We will adopt and compare these scenarios below and also comment on the possibility of multiple KN explosions in the two-step process.

In this paper we confront the \fe60 and \pu244 data with the SN and neutron-star merger nucleosynthesis models developed in~\cite{Wang2021} and explore the prospects of fresh measurements that might discriminate between possible scenarios, emphasizing the value of
analyzing lunar regolith samples returned to Earth by missions such as {\it Chang'e} and {\it Artemis}.

\section{Supernova and Kilonova Models of \fe60 and \pu244}
\label{sect:models}

The astrophysical origins of $r$-process elements including the actinides have been the subject of considerable debate~\citep{Cowan2021}. SNe may produce the {\em r}-process either via a neutrino-driven wind or 
in {magnetohydrodynamic (MHD)} jets,
but both
mechanisms struggle to make actinides; see~\cite{Wang2021} and references
therein. If SNe are confirmed as robust sources
of actinides such as \pu244, 
the available models must have major omissions. Neutron-star mergers that lead to KN explosions, on the other hand, have been observed to produce $r$-process species such as lanthanides~\citep{Abbott2017} and are expected to produce actinides~\citep{Cowan2021}, though the latter has yet to be confirmed observationally~\citep{Zhu}. 

In~\citet{Wang2021}, we constructed four SN and KN models to examine $r$-process radioisotope production.
Our two models featured a modified neutrino-driven wind \citep{arc11}
scenario {\it forced} to produce actinides, denoted by $\nu^*$ (SA), and {a high magnetic field} MHD SN model \citep{mhd}, denoted by SB, both with $r$-process nucleosynthesis constrained using data on the metal-poor star HD160617 \citep{Ian}. For neutron-star mergers we explored 
two combinations of calculations of neutron-star merger dynamical ejecta \citep{Bovard} and a disk 
$\nu$-driven wind \citep{Just}, constrained to fit data on either HD160617 (KA) or the actinide-boost star J0954+5246 \citep{Erika} (KB). Details of these models are found in \cite{Wang2021}.

\begin{figure}
  \includegraphics[width=0.45\textwidth]{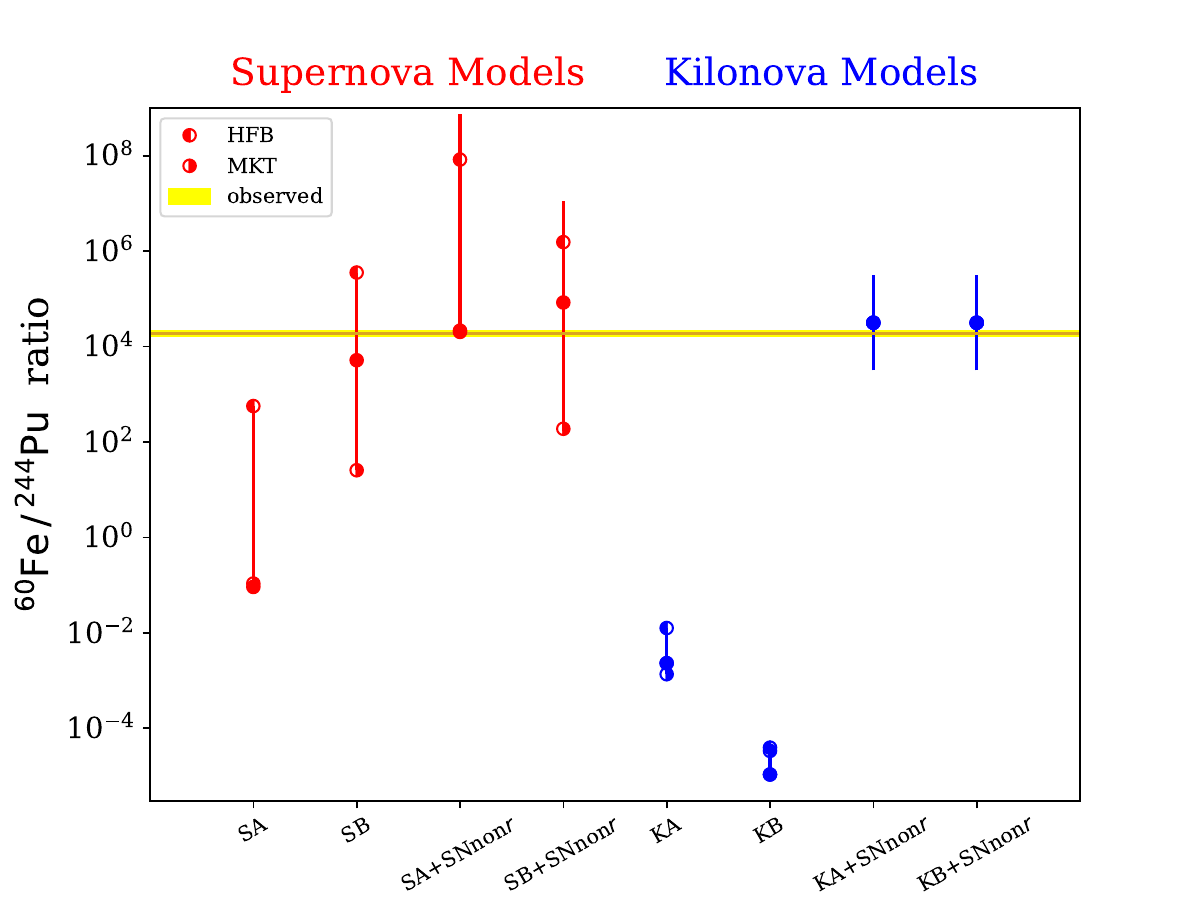}
	\caption{The \fe60/\pu244 abundance ratios calculated~\citep{Wang2021} in forced $\nu$ wind and MHD SN models (SA and SB)
	and in KN models (KA and KB). We present results for each model both without and including an additional non-$r$-process SN source of \fe60 at 100 pc; calculations are for events 3 Mya, but with a 10 Mya kilonova in the two-step KA/B+SNnon{\em r} models.
	The yellow band indicates the observed 
	\fe60/\pu244 ratio~\citep{Wallner2021} for 3 Mya (SN Plio).
The error bars indicating the impact {\em r}-process nuclear uncertainties discussed in Appendix~\ref{subsec:nucleosynthesis}: Filled symbols are the baseline FRDM+QRPA nuclear model, while the left (right) half-filled symbols are the HFB (MKT) models. 
}
\label{fig:FePu}
\end{figure}

The \fe60/\pu244 ratios for the four models of $r$-process production are compared to the data from~\cite{Wallner2021} in Fig.~\ref{fig:FePu}. In the SA and SB models, the \fe60 production is underestimated, as SNe produce \fe60 during hydrostatic burning phases and during the explosion \citep{Limongi2006,Suhkbold2016,Limongi2018,Curtis2019}, apart from any potential {\em r}-process contribution. Hence, in order to compare our models to the new data from~\cite{Wallner2021}, we need to consider sources of \fe60 within the SN event in addition to the $r$-process yields from~\cite{Wang2021}. Our procedure to do this appears in Appendix~\ref{subsec:nucleosynthesis}.

We see that
either of the SN models SA or SB could accommodate the (similar) \fe60/\pu244 ratios reported
by~\cite{Wallner2021} in the periods around 3 and 7~Mya. On the other hand, both the KN
models KA and KB predict much smaller \fe60/\pu244 ratios, even when the uncertainties are
taken into account. We therefore conclude that the \fe60 pulses and \pu244 detection {\em cannot be due to KN explosions alone}, at least as described by the models considered here.  

We consider now the data of~\cite{Wallner2021} on the \fe60 pulse from $\sim 3$~Mya.
The timing of this signal is consistent with that measured previously in \fe60 deposits
in deep-sea sediments and crusts~\citep{Knie1999, Knie2004, Fitoussi2008, Wallner2016, Ludwig2016,Wallner2020}, though this peak is somewhat broader. 
The observed amplitude of the pulse and its duration of $\gtrsim 1$~Myr are consistent with a model in which
\fe60 from an SN $100$~pc away is transported to Earth in dust grains via `pinball' trajectories
that are deflected and trapped by a magnetic field within the SN remnant
\citep{Fry2020,Ertel2022}.  The pulse width
indicated by the~\cite{Wallner2021} measurements could also reflect smearing in the crust they study.
Accordingly, we assume that this pulse was produced by a single SN.

In our SN models, we also make the economical assumption that the \pu244
from $\le 4.57$~Mya measured by~\cite{Wallner2021} is also associated with SN Plio 3 Myr ago.
We emphasize that observations with finer timing resolution would be needed to confirm this association, but note
that many of our comments below would apply also if it were due to two or more SNe.
As discussed above, the additional \fe60 peak discovered 
by~\cite{Wallner2021} -- see also Fig.~1 of~\cite{Fitoussi2008} --
is likely due to another SN that occurred $\sim 7$~Mya (SN Mio), also some $\sim 100$~pc away.
We assume that all the \pu244 from 4.57 to 9~Mya measured by~\cite{Wallner2021}
is associated with this SN explosion while emphasizing that observations with finer timing resolution 
would also be needed to confirm this association. Under this assumption, the \pu244/\fe60 ratios in the ejecta
of the two SNe $\sim 3$ and $\sim 7$~Mya are comparable within a factor of $\sim 2$ and indistinguishable in Fig.~\ref{fig:FePu}.

This is intriguing since simulations indicate that only very
specific types of SN can make much \pu244~\citep{Wang2021}, 
in which case seeing two of them looks like a remarkable coincidence.
If such an interpretation were correct, it would suggest not only that many or most SNe are $r$-process sites, 
but also that their production extends all the way to the actinides.
If this could be
established, standard $\nu$-driven wind and MHD models must have major omissions.
That said, actinide production is possible in the \textit{forced} neutrino wind and MHD models $\nu^{*}$ (SA) and SB discussed in~\cite{Wang2021}. 

Nevertheless, there are serious potential issues for scenarios with actinide production
in many or most SNe provided by measurements of the $r$-process abundances in metal-poor stars.
(a) It is known that $r$-process/Fe ratios (estimated using Eu/Fe as a proxy) vary wildly, with most stars showing low values
and only a minority showing high values \citep{Holmbeck+2020}. The obvious interpretation is that Fe and $r$-process production are decoupled.  
If SNe do indeed make the $r$-process, one possibility would be that (core-collapse) 
SN Fe production is highly variable. However, there are observational constraints on this from 
observations of SN light curves powered by $^{56}$Ni decay,
so it seems more likely that the variations in $r$-process/Fe ratios are due to variations in $r$-process production.  Another issue is that
(b) searches for $r$-process species in metal-poor dwarf galaxies
found them only in $\sim 10$\% \citep{Ji+2016}.  This strongly suggests that $r$-process events are much rarer
than SNe. An alternative hypothesis is that the $r$-process material is ejected preferentially from the dwarf galaxies, e.g.,
in jets, but in this case, jets would have to be features of most SNe, which is not
supported by observations.

Motivated by these considerations, we proposed in~\cite{Wang2021} that \pu244 signals could
arise via a two-step process in which material deposited previously in the interstellar medium (ISM) by an earlier
KN was then
swept up by subsequent SN explosions~\footnote{First estimates of the
possible amounts of swept-up \fe60 and \pu244 were given in~\cite{Ellis1996}.
We note that the transport of material from a KN can be complex.  For example, \citet{Amend2022} point out that if a KN explodes a few kiloparsec above the Galactic plane,  
the debris can take 1-10 Myr to fall on the disk.}. Estimates of the KN rate in the Galaxy are compatible with a KN explosion ${\cal O}(300)$~pc away that
occurred ${\cal O}(10)$~Mya. Accordingly, we also show in Fig.~\ref{fig:FePu} results from
models in which the debris from a KN explosion 10~Mya is mixed with the \fe60 of SN Plio 3~Mya.  This two-step model discussed in \cite{Wang2021} 
is consistent with the data shown in Fig.~\ref{fig:FePu},
and could explain naturally the similarity between the
\pu244/\fe60 ratios in the periods covering the two \fe60
pulses found by~\citet{Wallner2021}.

We calculate abundances in the two-step models via scaling with the \pu244 observations.  
We adopt the observed interstellar \pu244 flux $\Phi_{244} = 980 \ \rm atoms \ cm^{-2} \ Myr^{-1}$ as reported in \citet{Wallner2021}.  We then infer the other {\em r}-process fluxes using their production ratio to \pu244 and assuming the dust incorporation and survival fractions are the same as those for plutonium.  Finally, to compute fluence we assume that the flux results from stirring by SNe in the Local Bubble, which has continued since the KN explosion.
For the explosion times, we adopt 10, 20, and 50 Myr ago.

The two-step model in \citet{Wang2021} assumes a single KN explosion produces the \pu244 seen in the deep-sea deposits.  It is however possible that multiple such events could contribute to the signal since
the abundance deposition is driven by the stochastic nature of KN explosions \citep{Hotokezaka2015,Cote2021}. Models for {\em r}-process enrichment show large fluctuations on timescales below the recurrence time for a given location, which is estimated as $\sim 100-200 \ \rm Myr$ for turbulent diffusive mixing \citep{Hotokezaka2015,Paz2020}. This timescale makes multiple events unlikely in the $\sim 20 \ \rm Myr$ window available to Fe-Mn cursts, but not impossible.

We consider the case of multiple KNe in Appendix \ref{sect:manyKN}.  There we show that the signal from several events is a weighted average of signals from the individual events. Below we present KN model results for different explosion times spanning the plausible range, and so multiple events in this window would appear as signals bracketed by the cases we show.

\section{Deep-sea measurements}
\label{sect:sea}

Figure~\ref{fig:FePu} suggests that the current \fe60 and \pu244 data on their own are insufficient to discriminate between the SN-only estimates and the two-step KN/SN scenario, so we
consider also the possibility of observing additional SN pulses in deep-sea deposits
from $> 7$~Mya. A model of the Local Bubble and \fe60 transport proposed in~\cite{Breit2016, Schulreich2017} postulates 14-20 SNe in the Scorpius-Centaurus (Sco-Cen) stellar association within 300~pc over the past 13~My, among which might be progenitors for the two observed \fe60 pulses. 
\cite{Wallner2021} reported the results of \fe60 searches extending over the past 10~Myr, finding that the signal-to-background ratio for \fe60 falls to around unity for deposits from between 7 and 10~Mya. The relatively short \fe60 half-life of 2.6~My would make searches for earlier \fe60 pulses even more challenging.
On the other hand, indirect evidence for earlier SNe could come from pulses of swept-up \pu244 in earlier deep-sea deposits, in view of its much longer half-life $\sim 80$~My. 
\cite{Wallner2015} reported the results of a search for \pu244 extending over the past 25 Myr, finding one event from $> 12$~Mya. This event might just be background, but if not, it would correspond to a rate of deposition similar to that between 5 and 12~Mya. 
The more sensitive \pu244 results of~\cite{Wallner2021} extend back to 9~Mya, and it would clearly be interesting to extend the search for an earlier \pu244 signal and any possible time structure.

\section{Additional radioisotopes of interest}
\label{sect:additional}

Another, potentially more powerful, way to distinguish between the possible production scenarios is to look for other radioisotopes present alongside the \fe60 and \pu244.
Any $r$-process mechanism that produces \pu244 also produces many other radioisotopes,
not only other actinides such as \u236, \np237, and \cm247, but also many other radioisotopes
with masses intermediate between \pu244 and \fe60. Hence, their abundances would in general
exhibit pulses coincident with the two SN \fe60 pulses, whether the
$r$-process location is a recent, nearby SN or an earlier, more distant KN. However,
the relative abundances of the peaks of different $r$-process isotopes would be affected
by their lifetimes, which would help distinguish scenarios
in which the $r$-process occurred at different times in the past. 

We have calculated
the relative abundances of live $r$-process radioisotopes produced by the {\it forced} 
$\nu$-wind (SA) and MHD (SB) models
for SN Plio and Mio that occurred 3~Mya and 7~Mya discussed above, as well as two scenarios with a KN explosion 10 or 20~Mya.  
If the \pu244 measured in~\cite{Wallner2021} was produced by such an SN, one could hope to see accompanying signals of $r$-process production of the radioisotopes \zr93, \pd107, \i129, \cs135, \hf182, \u236, and possibly \np237 and \cm247, as listed in Table~\ref{tab:ratios-3and7}.
The first four columns of Table~\ref{tab:ratios-3and7} compare the yields of live $r$-process radioisotopes predicted by SN models SA and
SB for SNe that exploded 3 and 7 Mya. We emphasize that the \fe60/\pu244 ratios given in this table are only for \fe60
produced via the $r$-process and that we would expect these SNe to produce much more \fe60 via the standard neutron-capture mechanism. 
The last four columns in Table~\ref{tab:ratios-3and7} show results from calculations of 
$r$-process isotope production in KN models KA and KB,
assuming an event 10 or 20~Mya, bracketing the formation of the Local Bubble. In this case, many of the
shorter-lived radioisotopes that could have been detectable in the SN scenario
would have decayed away, and we find that if the measured \pu244 was produced by a KN 10 or 20~Mya, 
the best prospect for detection (with the biggest radioisotope ratio to \pu244) is for \i129.

\begin{table*}[htb!]
    \centering
    \caption{{\em r}-process isotope ratios in {\em forced} SN models for explosions 3/7 Mya,
    corresponding to the known \fe60 pulses, and
    in KN models for explosions 10/20~Mya, 
    bracketing the formation of the Local Bubble.}
    \label{tab:ratios-3and7}
\vspace{0.5cm}
\begin{tabular}{|c|cc|cc||cc|cc|}
    \hline \hline
     & \multicolumn{4}{c||}{Supernova Models} & \multicolumn{4}{c|}{Kilonova Models}\\
        \hline  
  Radioisotope   & \multicolumn{2}{c|}{SA} & \multicolumn{2}{c||}{SB} & \multicolumn{2}{c|}{KA} & \multicolumn{2}{c|}{KB} \\    
   Ratio  & 3~Mya & 7~Mya & 3~Mya & 7~Mya & 10~Mya & 20~Mya & 10~Mya & 20~Mya \\
    \hline  
    \fe60/\pu244 &$9.2\times 10^{-2}$ & $3.2\times10^{-2}$  & $5.3\times 10^{3}$ & $1.8\times10^{3}$ & $3.7 \times 10^{-4}$ & $3.0 \times 10^{-5}$ & $1.7 \times 10^{-6}$ & $1.4 \times 10^{-7}$\\
    
     \zr93/\pu244 &5.2 & 0.93  & $8.2\times 10^{4}$ & $1.5\times10^{4}$ & 0.24 & $3.6 \times 10^{-3}$ & $7.7 \times10^{-3}$ & $1.2 \times 10^{-4}$\\
      
    \pd107/\pu244 &52 & 35  & $1.3\times 10^{5}$ & $8.6\times10^{4}$ & 3.7 & 1.4 & 0.34 & 0.13 \\
    
    \i129/\pu244 &$3.2\times 10^{2}$ & $2.8\times10^{2}$  &$1.7\times 10^{6}$ & $1.5\times 10^{6}$ & 69 & 49 & 14 & 10 \\

    \cs135/\pu244 &5.4 & 0.68  &$1.2\times 10^{5}$ & $1.5\times10^{4}$ & $8.7 \times10^{-3}$ & $5.5 \times10^{-5}$& $3.7 \times10^{-2}$& $2.4 \times10^{-4}$\\

    \hf182/\pu244 &3.1 & 2.3 & $4.4\times 10^{3}$ & $3.3\times10^{3}$ & 0.43 & 0.22& $6.9 \times10^{-2}$ & $3.5 \times10^{-2}$\\
    \u236/\pu244 &1.8 & 1.7 &9.5 & 8.7 & 1.8 & 1.5 & 1.0 & 0.92\\
    \np237/\pu244 &0.66 & 0.18 &1.6 & 0.43 & $8.2 \times10^{-2}$ & $3.7 \times10^{-3}$ & $5.6 \times10^{-2}$ & $2.5 \times10^{-3}$ \\
    \cm247/\pu244 &0.50 & 0.43 &0.45 & 0.39 & 0.38 & 0.27 & 0.35 & 0.25 \\
    \hline \hline
    \end{tabular}
\end{table*}

\begin{figure}
    \centering
   \includegraphics[width=0.5\textwidth]{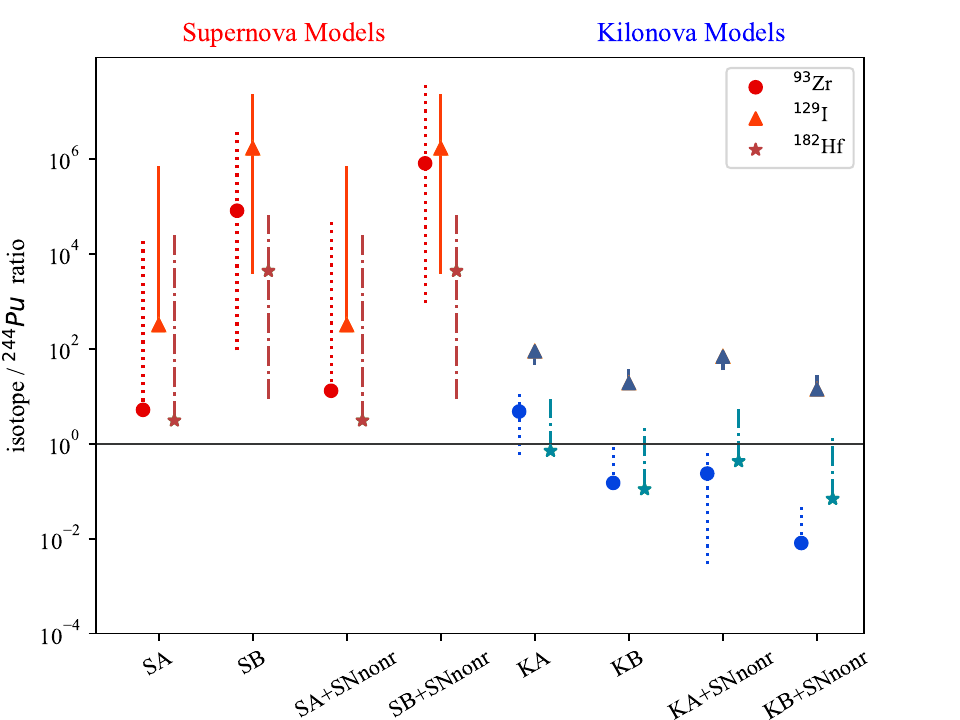}
\vspace{-5mm}
	\caption{Ratios to \pu244 of the selected live $r$-process radioisotopes \zr93, \i129 and \hf182, calculated in a similar way to Fig.~\ref{fig:FePu}. The vertical error bars indicate the impact of {\em r}-process nuclear uncertainties due to variations in masses and beta-decay rates as well as the fission yields, as discussed in Appendix~\ref{subsec:nucleosynthesis}.
	}
	\label{fig:plutonium2}
\end{figure}

Figure~\ref{fig:plutonium2} compares the total yields of selected live $r$-process radioisotopes predicted by our SN and KN models with direct deposition (one-step) as well as the two-step scenario (a 10~Mya KN plus a 3~Mya non-$r$-process SN), with similar calculations as in Fig.~\ref{fig:FePu}.
We highlight in Fig.~\ref{fig:plutonium2} representative isotopes in each of the three regions of the $r$-process abundance pattern, namely \zr93, \i129, and \hf182, which we now discuss in turn. \zr93 can be produced in an alpha-rich freezeout of mildly neutron-rich SN ejecta without an accompanying main $r$-process, so its detection could be a probe of this additional nucleosynthetic source.
Here we take the non-{\em r}-process SN \zr93 yields to be $M_{\rm ej,93} \sim 10^{-7.5} M_\odot$ (see Appendix~\ref{subsec:nucleosynthesis}). As for \i129, our predictions show that it should be detectable alongside \pu244 in any scenario, with measurement of the ratio offering possibly the strongest discrimination between scenarios. Finally, we anticipate that \hf182 could be a clear marker of prompt SN production, as it is present in potentially detectable levels for the SA and SB models, but not in the two-step KN scenarios.

It is a common feature of all the SN and KN models studied above that the best prospects for
discovering a second live $r$-process radioisotope (in addition to \pu244) may be offered by \i129\footnote{Note that iodine is volatile, and thus forms dust less readily than 
refractory elements, which include most other {\em r}-process radioisotopes of interest. (We thank Toni Wallner for pointing this out.)
While the cosmic dust properties of iodine are not well known \citep{Lodders2023} and merit further study,
it is possible that the surviving dust fraction of \i129 will be lower than other {\em r}-process elements including \pu244.}.
The \i129/\pu244 ratio calculated in the models we have studied
ranges from ${\cal O}(10)$ in the KN models through ${\cal O}(100)$ in SN model SA to 
${\cal O}(10^6)$ in SN model SB, thereby offering the possibility of distinguishing between scenarios. 
An \i129/\pu244 ratio exceeding $10^5$ that is
coincident with either of the observed SN pulses would favor SN model SB, which
would also predict 
\hf182/\pu244 ratios $> 10^3$. On the
other hand, a \i129/\pu244 ratio between $10^3$ and 10 could be accommodated by
any of the models SA, KA, and KB. 
In this case, model SA suggests that
\hf182 might be present at levels similar to \pu244, whereas the KN
models predict smaller ratios for \hf182 
relative to \pu244, which are less likely to be detectable. Hence, detection of
\hf182 at a level similar to \pu244 would point strongly towards
an SN $r$-process origin.
Additionally, detection of any of the other $r$-process radioisotopes \zr93, \pd107, {\cs135}, \hf182, 
and \np237
would favor an SN origin for the \pu244. 
A \u236 signal is possible in both the SN and KN scenarios but may suffer from anthropogenic or natural backgrounds in which \u235 can capture a neutron. While \cm247 has no natural background, the \cm247/\pu244 ratio is similar for SN and KN models and thus does not offer discriminating power. However, this ratio is sensitive to the unknown nuclear physics in the neutron-rich actinide region \citep{Holmbeck2019,Lund2022}, thus future \cm247 measurements can be used to test nuclear physics inputs, such as nuclear masses, half-lives, and fission properties.

\section{Lunar regolith searches}
\label{sect:lunar}

Lunar regolith (soil) serves as a natural archive for material from nearby explosions that is complementary to terrestrial samples.
Advantages of lunar archives include the geological inactivity of the Moon, the lack of an atmosphere or oceans that can redistribute material, and the lack of anthropogenic disturbances or contamination.
On the other hand, the lunar surface is slowly reworked by meteoritic ``gardening'' \citep{Gault1974,Costello2018}, so signals from multiple events will be mixed.  Moreover, the lunar surface is unshielded from cosmic rays, which create a background of radioisotopes in the regolith \citep{Reedy1972,Vogt1990,Leya2001,Leya2021}.  Any SN or KN signal must stand out from this background in order to be detectable.
Here we focus on the detectability of live $r$-process radioisotopes on the Moon.

Lunar regolith samples come at a great cost, but
our inventory will soon increase dramatically
beyond the {\em Apollo} and {\em Luna} samples
that date back to the 1960s and 1970s.
The robotic {\it Chang'e-5} mission~\citep{Change5}
has recently delivered a $\sim 1.7 \ \rm kg$ sample from 
a location farther north than any prior landings, the {\it Chang'e-6} mission will land in the South Pole region of the Moon in~2024,
and the upcoming
crewed {\it Artemis} mission~\citep{Artemis}
will bring back $\sim 100 \ \rm kg$ of samples in the initial landing near the South Pole, with more planned thereafter.

We recall that the discovery of \fe60 in
several {\it Apollo} samples was reported in~\cite{Fimiani2016}. The data of~\cite{Wallner2021} suggest that this \fe60 is likely
to have originated from a combination of the \fe60 pulses from 3 and 7~Mya, mainly the
more recent pulse in view of its greater fluence and younger age.
Confirmation of this \fe60 signal, e.g., in the sample returned
recently by the {\it Chang'e-5} mission~\citep{Change5} or that from a future 
{\it Artemis} lunar landing mission~\citep{Artemis}
would require analyzing a modest sample of $\lesssim 100$~mg of lunar material~\footnote{Employing the regolith gardening model of~\cite{Costello2018, Costello2020}, it is  estimated in~\cite{Change5} that the depth $\Lambda$ at which the probability of at least one overturn is 99\% is $3.45 \times 10^{-5} \ t_{\rm yr}^{0.47} \ \rm m$,
where $t_{\rm yr}$ is the reworking time in yr. This leads to  $\Lambda = 3.8\ (5.7)$~cm for material deposited 3 (7)~Mya.
}.

The detection of \pu244 in the deep sea implies that a corresponding lunar signal must exist as well.  Any cosmic-ray background for 
 \pu244 must arise from the available \u235 and \u238 targets.
 These require multiple $p$ and $n$ captures, and even then the $\sim 5 \ \rm hr$ half-life of \pu243 effectively diverts any cosmogenic flow away from \pu244.
Cosmic rays can also destroy \pu244 by neutron capture, but this effect is negligible compared to decay: In Appendix~\ref{subsec:destruction},
we estimate $\Gamma_{244+n} \tau_{244} \sim 4 \times 10^{-8} \ll 1$.
We therefore conclude that
cosmic-ray processes do not build up any appreciable \pu244 background in the regolith
nor do they destroy it. \pu244 should thus be a particularly clean lunar target, as long as return samples are 
protected from terrestrial contamination.

In the deep-sea Fe-Mn crust, the relative 
abundance of \pu244 reported by~\cite{Wallner2021} is $\pu244/\fe60 \sim 5 \times 10^{-5}$. Since it is expected that uptake on the lunar surface is as efficient as in the deep-sea case, if not more, we anticipate sample sizes of similar mass ($\sim 10~\rm g$) would be needed to discover a \pu244 signal.

\begin{figure}[htb]
    \centering
    \vspace{-3mm}
    \includegraphics[width=0.49\textwidth]{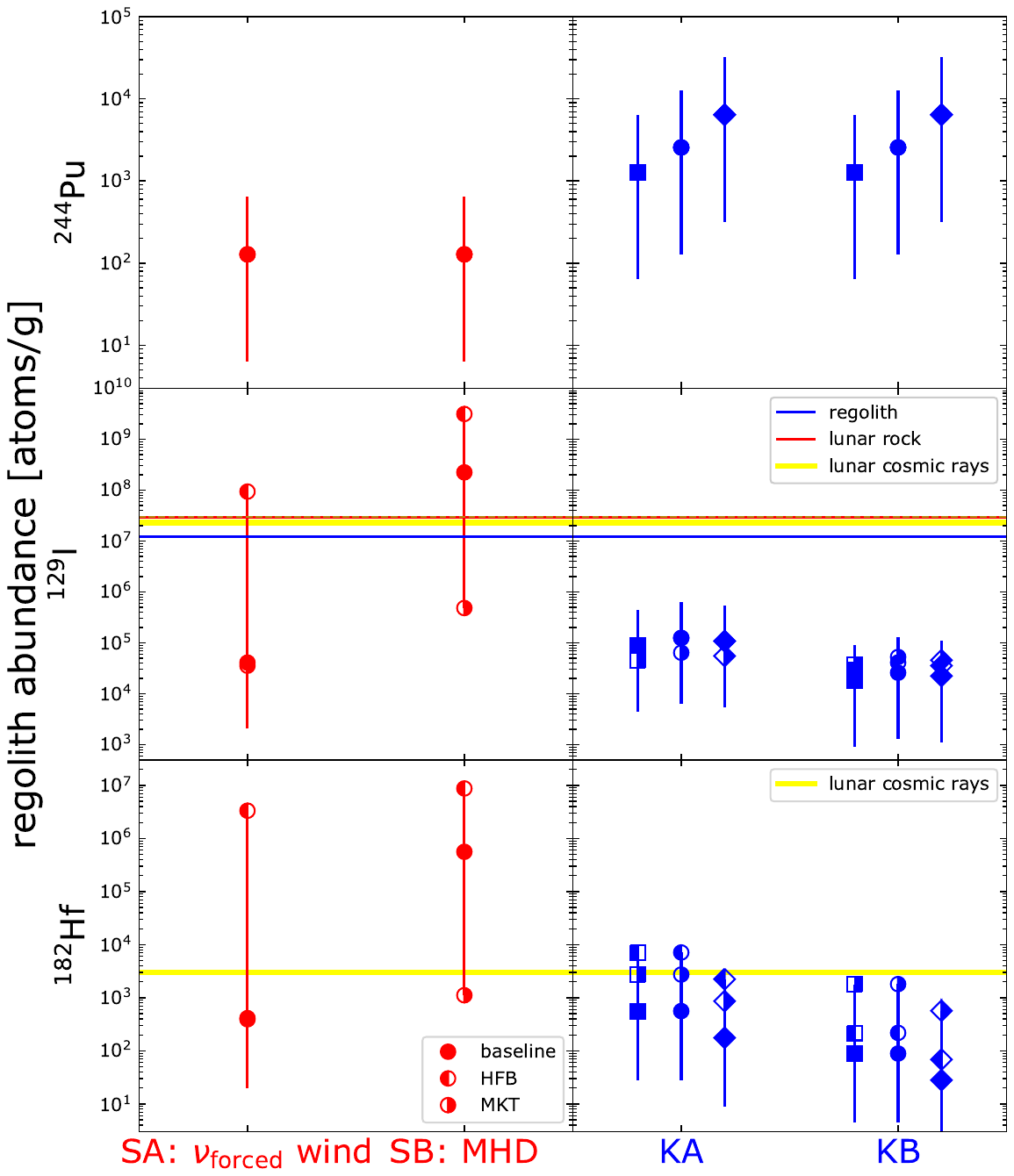}
    \caption{Expected abundances of select {\em r}-process radionuclides in lunar regolith, based on \pu244 deep-ocean measurements.
    Points show predictions for SN and KN models,
    with error bars indicating the impact 
    {\em r}-process nuclear uncertainties discussed in the Appendix~\ref{subsec:nucleosynthesis}: Filled symbols are the baseline FRDM+QRPA nuclear model, while the left (right) half-filled symbols are the HFB (MKT) models.
    For \i129, the horizontal blue and red lines show the
    levels measured in {\em Apollo} samples of lunar regolith \citep{Nishiizumi1989} and rock \citep{Nishiizumi1983},
    and the yellow line shows the calculated cosmic-ray background (for details of the cosmic-ray background estimations, see Appendix~\ref{subsec:production}). SN models are for an explosion 3 Myr ago; KN models show results for explosions 10, 20, and 50 Myr ago.}
    \label{fig:rpro-regolith}
\end{figure}

For other, lighter {\em r}-process radioisotopes,
the lunar regolith will contain a cosmogenic background that we must understand.  
Here we recommend the strategy that has successfully identified SN \fe60 in lunar regolith in the presence of a cosmic-ray background \citep{Fimiani2016}, as discussed in Appendix~\ref{subsec:distinguish}.  
A radioisotope $i$ with a potential astrophysical signal will also have a cosmogenic component,
so regolith abundances will sum the two, so we write the number per unit mass as $y_{\rm i}^{\rm obs} = y_i^\star + y_i^{\rm cr}$.  The cosmogenic background in each part of the sample will depend on the cosmic-ray flux $\Phi$ and the local target abundance: $y_i^{\rm cr} \propto \Phi y_{{\rm target}|i}$, where cosmic rays produce our species via ${\rm CR}+{\rm target} \rightarrow i$.  The target abundance is directly observable in the return sample, but cosmic-ray flux is not.  To infer the local cosmic-ray exposure reliably requires the measurement of a radioisotope species $j$ that is dominated by cosmic-ray production.  This approach was used to measure SN-produced \fe60 in the lunar regolith by also measuring \mn53 \citep{Fimiani2016}.  Such a two-radioisotope ``dyadic'' approach \citep{Koll2022} is elaborated in Appendix~\ref{subsec:distinguish}, which shows that a plot of the ratio $y_i^{\rm obs}/y_j^{\rm obs}$ as a function of the ratio of cosmic-ray targets $y_{{\rm target}|i}/y_{{\rm target}|j}$ falls on a line if species $i$ is cosmogenic only, while excursions above this line would indicate the presence of an additional extrasolar component.

While the {\em r}-process actinide radioisotopes \pu244 and \cm247 will not have an appreciable cosmic-ray background, a substantial background will exist for \u236 and \np237, which can be created from lunar \u235.  Cosmogenic production will also be an issue for all of the lighter species we consider.  Of the lighter species, \zr93, \pd107, and \cs135 have abundant cosmogenic target nuclei in the regolith, which leads to prohibitively large backgrounds.  We therefore focus on \i129 and \hf182 where the background is smaller.

Table \ref{tab:lunarobs} and Fig.~\ref{fig:rpro-regolith} summarize the prospects for lunar detection of \pu244, as  well as for \i129 and \hf182 including cosmogenic backgrounds estimated as described in Appendix~\ref{subsec:production}.  We also estimate the minimum number ${\cal N}_{i,\rm min}$ of atoms of each species needed for a confident accelerator mass spectrometry (AMS) measurement, based on present sensitivity levels and extraction efficiency.  From the needed number of atoms and the predicted number $y_i = n_i/\rho$ of signal atoms per gram, we estimate the sample mass $m_{i,\rm min} = {\cal N}_{i,\rm min}/y_i$ needed; for the SN case, this spans a large range, reflecting the large uncertainties in the predictions.  

The AMS sensitivity for \i129 together with large \i129/\pu244 predicted ratios
suggests that \i129 detection is already within the grasp of present techniques (if
care is taken to avoid contamination from anthropogenic sources); indeed, intriguing measurements already exist.
\i129 has already been detected in a Fe-Mn crust \citep{Ji2015ams},
showing a dropoff with depth consistent with a background source
such as natural uranium fission.
The Fe-Mn crust was not independently dated,
but the abundance levels appear inconsistent with the SB model while allowing room for some SA models
and the KA and KB models.
We note that~\cite{Nishiizumi1989} used AMS to measure 
\i129 in the lunar regolith, and \cite{Nishiizumi1983} measured it in 
lunar rock, finding very similar abundances.  Lunar rocks should not contain an SN or KN component so, at face value, these data seem to place an upper limit $n(\i129)/\rho \lesssim 3 \times 10^{7} \, \rm atoms \ {\rm g}^{-1}$ on any extrasolar signal.  This result is consistent with the indications from \i129 data in a Fe-Mn crust \citep{ji2015crusts} discussed in \cite{Wang2021}.  These measurements were not made with our research program in mind, but still, the apparent nondetection of \i129 puts pressure on the SN models, as seen in Fig.~\ref{fig:rpro-regolith}:  It would rule out the HFB nuclear model for the forced neutrino wind case, 
as well as the baseline and HFB nuclear models for the MHD SN. Future dedicated \i129 searches in the deep ocean and on the Moon would be of great interest.

Compared to \i129, current AMS capabilities offer less promise for detecting \hf182. As estimated in Table~\ref{tab:lunarobs}, 
several kilograms of material may be required to detect a signal due to the limitations on AMS sensitivities caused by the interference from $A = 182$ isobars. Due to the scarcity of available lunar material, higher sensitivity would be desirable to reduce the quantity that must be processed. Figure~\ref{fig:rpro-regolith} motivates this effort, as the cosmic-ray production of \hf182 is predicted to be closer to the KN predictions than that of \i129. These data foreshadow the power of new radioisotope measurements on both terrestrial and lunar samples, particularly \i129 and \hf182, which can probe the nature of the recent explosions and of the {\em r}-process generally.

Finally, we note that the {\em spatial} distribution of \fe60 and other radioisotopes on the Moon carries information about the supernova direction and dust propagation~\citep{Fry2016}.  Over the long timescale of dust deposition,  lunar rotation will average the deposition over longitude, but not latitude.  Consequently, the latitude distribution of radioisotopes probes the distribution of dust arrival directions.  If the dust arrives in a plane wave, there will be pronounced gradients that should be detectable, but the extant {\em Apollo} \fe60 data are too uncertain to test for such a gradient \citep{Fry2016}.  Future measurements of {\em Artemis} and {\em Chang'e} samples, particularly from landings near the lunar poles, could reveal the latitude distribution and thus give unique insight into the dust propagation.  The results we show in Fig.~\ref{fig:rpro-regolith} and Table \ref{tab:lunarobs} are for the latitude where a plane-wave flux arrives vertically; this will be modulated by the arrival direction distribution.

\begin{table*}[htb!]
    \centering
    \caption{Lunar Regolith {\em r}-Process Radioisotopes From Near-Earth Explosions}
    \begin{tabular}{c|ccc|cc}
    \hline\hline
& Cosmic-Ray &  AMS Sensitivity & & \multicolumn{2}{c}{Sample Mass (g)} \\
        Isotope & Targets & (atoms) &  {Background (atoms g$^{-1}$)} &  SN & KN \\
    \hline 
    \i129 &  Te, Ba, La & $10^5$ & $10^7$ & $10^{-1}-10^3$ & 1--10 \\
    \hf182 & \iso{W}{183}, \iso{W}{184}, \iso{W}{186} & $10^7$ & $3 \times 10^3$ & $2-5\times10^5$ & $3 \times 10^3-10^6$ \\
    \pu244 & -- & $10^2$ & --  & \multicolumn{2}{c}{10} \\
    \hline \hline
    \end{tabular}
    \label{tab:lunarobs}
\end{table*}

\section{Cosmic-Ray Measurements}
\label{sect:cosmicray}

There are also possible signatures of nearby explosions in cosmic rays, including anomalies in positron and antiproton fluxes \citep{Kachelriess2018}, and also in the ion composition.
In particular, \fe60 is seen in cosmic rays \citep{Binns2016}, and anomalies in elemental iron fluxes at low energies also suggest a perturbation due to a recent event \citep{Boschini2021}.  

A nearby $r$-process event would mainly produce stable isotopes, which would be difficult to identify in deposits on the Earth or Moon, but might be detectable among the cosmic rays. 
Searches for heavy elements in cosmic rays have led to several intriguing new results. The Cosmic Ray Isotope Spectrometer (CRIS) recently reported data on elements with atomic number $29\leq Z \leq 38$ and found lower abundances of $r$-process species than would be expected if SNe were their source~\citep{Binns:2022qmw}. 
SuperTIGER has recently reported cosmic-ray abundances for heavier species as well \citep{Walsh,WalshThesis}.
The $42 \le Z \le 54$ data from SuperTIGER shows anomalously high abundances, i.e., exceeding the levels of a mix of 80\% solar-system-like material with a 20\% admixture of SN winds and ejecta that fit lower-mass cosmic-ray species. Some of these anomalous elements are produced mainly by the $r$-process, though with admixtures of $s$-process production. 
Intriguingly, the dominantly {\em s}-process species barium is not as enhanced as the other high-$Z$ elements, which may be circumstantial support for the hypothesis of a 
nearby $r$-process site~\citep{WalshThesis}. 

Cosmic-ray measurements of other {\em r}-process species such as actinides could shed light on the nucleosynthesis pattern and thus the {\em r}-process source. If isotopic measurement were possible in future cosmic-ray experiments, observation of \hf182 would be particularly interesting, since it should not have significant contamination from the spallation of neighboring nuclides.  
\i129 and \cs135 would also be of interest, though contamination from spallation of Te, Xe, and Ba may be an issue.

There is compelling evidence that SN remnants give rise to the bulk of Galactic cosmic rays, via diffusive shock acceleration \citep{Ackermann2013}.   But this same process should act in the relativistic blast waves generated in the gamma-ray bursts and KNe following neutron-star mergers, which therefore should also accelerate cosmic rays \citep{Waxman1995}.  Thus, in both the KN and SN scenarios, we expect cosmic rays to include freshly synthesized {\em r}-process material.  The elemental ratios for each scenario are shaped by the prompt production as well as the spallation effects that occur as the material is ejected from the SN/KN and propagates through the ISM \citep{Wang2019}. Whether such ratios can offer an opportunity to discriminate among scenarios requires further study.

\section{A Strategy to Determine the Origin of Near-Earth \pu244}
\label{sect:strategy}

By combining terrestrial and lunar measurements, we can hope to have information at least on the broad time history and overall fluence of \fe60 as produced in multiple SNe and separately the time history and fluence of \pu244, probably \i129, and possibly \hf182.  This will give insight into the origin of the {\em r}-process signal in radioisotopes and thus open a new window into the astrophysics site of the {\em r}-process.
We distinguish three possible cases.
\begin{itemize}
    \item
    No pulse coincidence: If \pu244 has a time history distinct from the \fe60 pulses, this would point to a different origin, likely a KN that seeded the Local Bubble.  Here the transport to Earth would not be coincident with the SN blasts and imply that the {\em r}-process-bearing dust moves independently in the bubble interior (as in our simple version of the two-step model).
    The radioisotope ratios $r/\pu244$ should follow the KN predictions, and we expect signals of \pu244 and longer-lived {\em r}-process radioisotopes going back to earlier times, tracing a prolonged flux back to the origin of the Local Bubble.
    
        \item
        One pulse coincidence: If \pu244 and other {\em r}-process radioisotopes trace one (and only one) of the \fe60 pulses, this points to a supernova origin of the {\em r}-process, and indeed an SN origin of {\em r}-process actinides.  This would have major implications for SN physics and for Galactic chemical evolution. In this case, the $\pu244/\fe60$ ratio would probe SN actinide production.
    
    \item
    Two pulse coincidences: If \pu244 traces {\em both} \fe60 pulses, this would either (a) require that both SNe produced {\em r}-process actinides, 
    which would challenge the prevailing SN nucleosynthesis models,
    or that (b) the Local Bubble was seeded with recent {\em r}-process events whose radioisotopes were later delivered by the SN blasts.  This would point to a KN origin for {\em r}-process actinides.  As we have shown, radioisotope ratios, $r/\pu244$, can distinguish these cases.

    High measured $r/\pu244$ ratios would indicate SN origins as in (a).  This would be quite unexpected, as it would require that the Local Bubble harbored at least two {\em r}-process SNe, and these were among the nearest events, suggesting that SNe produce the {\em r}-process much more commonly than has been thought.  The $\fe60/\pu244$ and $\i129:\hf182:\pu244$ ratios would probe the uniformity of the {\em r}-process synthesis in these two explosions.  
    
   Lower $r/\pu244$ ratios would point to a KN scenario as in (b).  The two SNe could have different \fe60 yields, so the two $r/\fe60$ ratios could vary (although \cite{Wallner2021} find that \fe60/\pu244 is consistent with a constant ratio).  But the common origin of the {\em r}-process species means that the ratios among the {\em r}-process species should be the same (within errors) for the two pulses.  Thus, consistency of the $\i129: \hf182:\pu244$ ratios would provide a check on this scenario. Note also that these ratios should be similar in the no-pulse scenario above, and here too we predict signals of \pu244 at earlier times prior to SN Mio 7 Myr ago; both of these are hallmarks of the KN origin.  Thus, the difference in the {\em r}-process deposition history between the zero and two-pulse pictures would serve as a probe of transport of KN debris within the Local Bubble.

\end{itemize}
In all cases, the $r/\pu244$ ratios probe in detail the nucleosynthesis and dust formation of these species.  If the dust formation properties are similar, then these species probe the {\em r}-process pattern in the source identified by the time history results.  In particular, \i129 probes the production of the second {\em r}-process peak, while \hf182 probes the base of the third peak. Observational data linking the second and third {\em r}-process peaks from the same source are limited, in large part because second-peak elements such as Te can only be spectroscopically identified in the UV.  There are hints in these data that Te production may be more strongly correlated with the first {\em r}-process peak than the third \citep{Roederer+2022}. A coincident detection of \i129 and \hf182, therefore, would add to this limited body of data and offer an independent test of this intriguing suggestion.~\footnote{Future searches for radioactive decay lines from $r$-process radioisotopes by next-generation MeV gamma-ray observatories such as COSI (\url{https://cosi.ssl.berkeley.edu}), AMEGO (\url{https://asd.gsfc.nasa.gov/amego/}), and MeVGRO (\url{https://indico.icranet.org/event/1/contributions/777/}) may also provide  information on near-Earth nucleosynthesis events \citep[e.g.,][]{Wang2020}.}

It is worth noting that, as we have seen, the one reported measurement of \i129 in lunar regolith \citep{Nishiizumi1989} apparently rules out much of the SN parameter space.  This is also apparently the case for the one reported \i129 measurement in deep-ocean Fe-Mn crusts \citep{ji2015crusts,Wang2021}, though the crust has not been independently dated.  These tantalizing past results were not made with our research program in mind, and follow-up measurements are in order to confirm these hints and to better illuminate cosmogenic backgrounds.  But the results in hand show that the experimental sensitivity to \i129 already within reach can enable dedicated measurements to shed new light on the {\em r}-process.

\section{Discussion: Connection with Early Solar System Radioactivities}

There is a close parallel between the study of live radioisotopes from recent nearby explosions and the study of extinct early solar system (ESS) radioactivities found in meteorites \citep[see, e.g.,][]{Huss2009,Dauphas2011,Lugaro2018,Davis2022}. 
Many of the central issues are identical.  In both cases, a nearby explosion leaves a signature in the form of a small isotope anomaly that is detectable with the aid of sophisticated laboratory techniques.  
Interpreting the detections requires a model for radioisotope nucleosynthesis, transport, and sequestration in the final sample, and the implications are broad, including probes of nucleosynthesis and of SN interactions with their environment.

Of course, the two situations also have important differences, which tend to make their strengths and weaknesses complementary.  In the case of recent nearby explosions, the signals are resolved in time, so that multiple events can be distinguished.  Also, the solar system remains in the environment that hosted the explosions, so that local astronomical observations probe its properties. Moreover, the conditions of the solar system at the time of radioisotope injection are surely similar to today and thus fairly well understood. However, the recent signals in geological samples are smaller than ESS anomalies, requiring AMS, and thus only accessible in a small number of laboratories.  In the case of ESS radioisotopes, the relatively larger isotope ratios allow for more measurements and thus a wide array of species can be searched for and found, bringing to bear the power of multiple isotope ratios.  On the other hand, the nature of the pre-solar nebula, and its larger environment, are not directly observable and thus are more uncertain.

ESS studies have found evidence for {\em r}-process species including not only \i129 and \cm247
but also \pu244 \citep[found via its spontaneous fission products in the form of xenon isotope anomalies in][]{Turner2007}.
 The ESS radioisotope inventory thus 
 includes both \fe60 and \pu244, in close parallel to the deep-sea results.
Because of these similarities, progress in studies of recent and ESS nearby events can and should be linked.  Indeed, as we have seen, the detection of \pu244 in both cases places important and complementary constraints on the {\em r}-process.  And both  studies have a close interplay with models for radioisotope production and mixing on local and Galactic scales \citep[e.g.,][]{Hotokezaka2015,Fujimoto2018,Lugaro2018,Cote2019ChemEv,Cote2021,Fujimoto2021}.
We urge that these connections be explored more deeply in the future.

\section{Conclusion:  Unearthing the Origin of {\em r}-Process Radioisotopes}
\label{sect:conclusion}

The combination of lunar and deep-sea probes of radioisotopes 
are complementary and may reveal their origin.  
\begin{itemize}
    \item Deep-sea crusts characterize the broad time history, e.g., the number of \fe60 pulses.  Future \pu244 measurements with better time resolution can test the coincidence with \fe60 and probe for events $\gtrsim 8 \ \rm Myr$ ago.  Searches for other {\em r}-process species test the SN and KN models, and the \i129 radioisotope has already been found in one crust.
    
    \item Deep-sea sediments give high-resolution time history, determining the timescales and, with improved sensitivity, the time profiles of the pulses.
    
    \item Lunar measurements avoid anthropogenic contamination, and so would offer important confirmation of the \pu244 detections, and are complementary to deep-sea crusts as sites for other {\em r}-process species that can discriminate between the SN and KN scenarios.   But regolith abundances require careful accounting for cosmic-ray production via measurements of target abundances and accompanying cosmogenic-dominated radioisotopes that encode exposure doses. 
    
    \item Lunar surface density measurements give the total fluence at Earth, summing over all events and avoiding uncertainties associated with terrestrial uptake factors.
    
    \item The lunar depth profile combined with a gardening model can give timing information or, vice versa, its combination with sediment data would give a new constraint on gardening.  It is possible that the depth profile might give indications of multiple events, but local stochastic variations might make this difficult.
    
    \item Comparing the \fe60/\pu244
    and $r$/\pu244 in deep-sea crusts and sediments probes uptake and helps determine the fluence independently of the lunar measurements.
    
    \item  Comparing the lunar \fe60/\pu244
    and $r$/\pu244 with those in crusts probes uptake, and comparison with sediment data probes nonuniformity of terrestrial deposition and possibly lunar impact losses and thus impactor velocity.

\end{itemize}

The models that indicate $> 10$ SNe formed the Local Bubble~\citep{Breit2016, Schulreich2017} suggest that there may be pulses of radioisotopes still earlier than SN Mio at 7 Myr ago.  This regime is probably beyond the reach of \fe60 due to decay, but fortunately, all of \i129, \hf182, and \pu244 can probe to 10 Myr ago and earlier, so any early {\em r}-process signal would remain.  To search for a correlation with SNe would require a long-lived radioisotope likely made in core collapse; \sm146 could be a candidate. 
The implications of such events are broad, with consequences for the heliosphere
\citep{Fields2008,Miller2022Helio,Miller2022Near}
and possibly for the biosphere \citep{Ruderman1974, Hartmann2002,Melott2004,Melott2017}.
Other astrophysical observables include searches for nearby neutron stars whose location and kinematics point to a Local Bubble origin \citep{Tetzlaff2010,Neuhauser2020,Zheng2022,Lin2022}.

The studies recommended in this work capitalize on recent technical advances, including expanded capabilities at AMS facilities to perform high-resolution measurements and the promise of next-generation radioactive beam facilities to reduce the considerable nuclear physics uncertainties of nucleosynthetic yield estimates \citep{Mumpower+16,Horowitz+2019,Schatz+2022}. We look forward to future studies of the deep-ocean crust---including more data on the 7 Mya SN Mio and earlier samples---as well as further data from sediments and the Antarctic snow. As we have argued here, lunar samples offer a unique complement to terrestrial sources and hold the promise to distinguish among {\em r}-process production scenarios, for which we anticipate results from the sample return missions of {\it Chang’e}
and {\it Artemis}.

\begin{acknowledgments}

We are grateful for illuminating discussions with Terri Brandt and Brian Rauch about SuperTIGER and cosmic rays, and to Toni Wallner and Dominik Koll for
discussions of their work.
X.W., R.S., and B.D.F. acknowledge many useful
discussions in the INT-21-3 workshop on cosmic radioisotopes sponsored by the Institute for Nuclear Theory.
The work of X.W. was supported by the US National Science Foundation (NSF) under grant Nos. PHY-1630782 and PHY-2020275 for the Network for Neutrinos, Nuclear Astrophysics, and Symmetries (N3AS) and by the Heising-Simons Foundation under award 00F1C7.
The work of A.M.C. was supported by the US Nuclear Regulatory Commission Award 31310019M0037 
and the National Science Foundation under grant No. PHY-2011890.
The work of J.E. was supported partly by the United Kingdom STFC Grant ST/T000759/1 
and partly by the Estonian Research Council via a Mobilitas Pluss grant. 
The work of A.F.E.~and B.D.F.~was supported by
 the NSF under grant No. AST-2108589.
 The work of J.A.M. was supported by the Future Investigators in NASA Earth and Space Science and Technology (FINESST) program under grant No. 80NSSC20K1515.
The work of R.S. was supported by N3AS as well as the US Department of Energy under contract Nos. DE-FG02-95-ER40934 and LA22-ML-DE-FOA-2440. R.S. also acknowledges the Aspen Center for Physics, which is supported by NSF grant PHY-2210452.

\end{acknowledgments}

\noindent
\software{\\
 Matplotlib \citep{matplotlib}, \\ 
 Numpy \citep{numpy1, numpy2}, \\ 
 Portable Routines for Integrated nucleoSynthesis Modeling (PRISM)~\citep{Mumpower2018,Sprouse2020}.}

\appendix

\section{Nucleosynthetic yield estimates and predicted ratios of radioisotopes}
\label{subsec:nucleosynthesis}

The measured ratios of $\fe60/\pu244$  and other radioisotopes 
reflect the interstellar fluence ratios for these species.  For species of SN origin such as \fe60, we use the usual expression \citep{Ellis1996,Fry2015}
for fluence (i.e., the time-integrated flux), 
${\cal F}_{60} = f_{\rm Fe} M_{\rm ej,60}e^{-t/\tau_{60}}/(4\pi A_{60} m_{\rm u} r_{\rm SN}^2)$,
in the case of \fe60 and similarly for other species.
Here the dust fraction measures the portion of the \fe60 ejecta arriving in grains. 
For the fluence of an individual species, one must specify the dust fraction as well as the distance,
but {\em ratios} of SN species are independent of distance;
they depend only on the dust fractions and yields
as well as basic nuclear properties:
\begin{equation}
    \frac{{\cal F}_i}{{\cal F}_j} = \frac{A_j}{A_i} 
    \frac{f_i}{f_j} 
    \frac{M_{{\rm ej},i}}{M_{{\rm ej},j}}
    \exp\left[ - t_i (\tau_i^{-1}-\tau_j^{-1}) \right]
\end{equation}
Throughout we take $f_i/f_j = 1$, i.e., we assume the same dust incorporation efficiency for all species of interest.  Thus, for the SN (one-step) case, the ratios depend only on the yields. 
For the KN case, we use the two-step model from \cite{Wang2021}, with a KN distance of 1000 pc.

Our {\em r}-process nucleosynthesis calculations are made using the nuclear reaction network code 
Portable Routines for Integrated nucleoSynthesis Modeling (PRISM)~\citep{Mumpower2018,Sprouse2020},
as implemented in~\cite{Wang2021}, {with baseline nuclear data from~\cite{FRDM2012} and \cite{MollerQRPA} (FRDM+QRPA), and variations in the masses~\citep{HFB17PRL} (HFB), $\beta$-decay rates~\citep{MKT} (MKT), and fission yields~\citep{KT}.

To combine the {\em r}-process \fe60 with the ordinary SN production requires that we specify the \fe60 mass yields for both ordinary and {\em r}-process synthesis. Turning first to SN yields, 
gamma-ray line observations provide an observational indication of the
mean \fe60 yield.  Given a Galactic steady-state \fe60 mass $M_{60,\rm ss} = 2.85 \, M_\odot$ \citep{Diehl2021}, and a core-collapse SN rate $R_{\rm SN} = 1.7 \, \rm events/century$
\citep{Rozwadowska2021}, the mean \fe60 yield is
$M_{\rm ej,60} = M_{60,\rm ss}/\tau_{60} R_{\rm SN}
= 4.5 \times 10^{-5} \, M_\odot$, where $\tau_{60}$ is the \fe60 lifetime; the
uncertainty in this mean yield is at least a factor of 2.
However, nucleosynthesis calculations suggest that \fe60 yields from individual SNe
span a wide range, varying sensitively and nonmonotonically with 
progenitor mass.
Yields in ref.~\cite{Suhkbold2016} lie span $(4 \times 10^{-6},3 \times 10^{-4}) M_\odot$,
a range that includes the results from calculations of \cite{Limongi2018} and model w of \cite{Curtis2019}.
We thus adopt an `ordinary' \fe60 yield of $M_{\rm ej,60} =10^{-4.5 \pm 1} M_\odot$. The models labeled `SA+SNnonr' and `SB+SNnonr' in Fig.~\ref{fig:FePu} include this \fe60 yield. Similarly, the yields of \zr93 from ordinary (non-{\em r}-process) SN explosions are in the range $(1.4 \times 10^{-9}, 2.4 \times 10^{-7}) M_\odot$ \citep{Limongi2018, Curtis2019}, thus an `ordinary' \zr93 yield of $M_{\rm ej,93} =10^{-7.5 \pm 1} M_\odot$ is included in the models labeled `SA+SNnonr' and `SB+SNnonr' in Fig.~\ref{fig:plutonium2}.

For the {\em r}-process mass yields, we adopt \citet{Wang2021}'s estimate for both the SN and KN models: $M_{{\rm ej},r}({\rm SA, SB, KA,KB})= (1.37\times 10^{-5},3.0\times 10^{-2},1.76\times 10^{-2},7.00\times 10^{-3})M_\odot$. The \pu244 and \fe60 {\em r}-process mass yields are then $M_{\rm ej,244}=A_{\rm 244} Y_{244} M_{{\rm ej},r}$, $M_{\rm ej,60}=A_{\rm 60} Y_{60} M_{{\rm ej},r}$.
We add the two \fe60 yields for the SN models .

For \pu244, we use the observed interstellar \pu244 flux \citep{Wallner2021} to determine
the fluence
$F_{244} = \Phi_{244}^{\rm interstellar} \ \Delta t$,
where $\Delta t$ is the time span of the measurement interval.
The {\em r}-process contribution to \fe60 then follows as
$F_{60,r} = f_{\rm Fe}/f_{\rm Pu} \ (\fe60/\pu244)_r F_{244}$,
where $(\fe60/\pu244)_r$ is the model abundance ratio by number calculated above.
The other {\em r}-process species have fluence ratios just given by their production number ratios, again assuming dust fractions $f_i/f_{\rm Pu}=1$.}

\section{Radioisotope Signals for Multiple Kilonovae}
\label{sect:manyKN}





In the KN scenario, we have considered the case of a single KN explosion that
enriches the proto-Local Bubble with {\em r}-process radioisotopes.  Here we consider the possibility of multiple nearby KNe contributing the the observed signal. For simplicity--and based on expectations from KN rates--we will consider the case of two KNe, but it will be clear how the result generalizes.

The predicted radioisotope signatures are all linked to the observed \pu244 flux, and 
the predictions for other isotopes can be expressed via their ratios to \pu244.
Thus, we consider a radioisotope species $A_i$ in the case of two KNe.  We would measure the ratio given by the combined depositions (fluences) from the two events:
\begin{eqnarray}
\pfrac{A_i}{\pu244}_{\rm obs} & = &  
\frac{\Phi_{i,1} e^{-t_1/\tau_i} \Delta t_1 + \Phi_{i,2} e^{-t_2/\tau_i} \Delta t_2}{\Phi_{244,1} e^{-t_1/\tau_{244}} \Delta t_1 + \Phi_{244,2} e^{-t_2/\tau_244} \Delta t_2} \\
& = & w_1 \pfrac{A_i}{\pu244}_{1} + w_2 \pfrac{A_i}{\pu244}_{2}  \ \ .
\label{eq:2KNe}
\end{eqnarray}
We see that the observed ratio is a weighted sum of the ratio
$(A_i/\pu244)_1 = (\Phi_i/\Phi_{244})_1$ for KN 1 alone and the corresponding ratio for the second event.
The weightings depend on the \pu244 fluxes, durations, and times of the two
events--that is, the weightings are the decayed plutonium fluences.  Namely, we have
\begin{eqnarray}
w_1 & = & \frac{(\Phi_{244,1}/\Phi_{244,2}) (\Delta t_1/\Delta t_2) e^{-(t_1-t_2)/\tau_{244}}}{1+(\Phi_{244,1}/\Phi_{244,2}) (\Delta t_1/\Delta t_2) e^{-(t_1-t_2)/\tau_{244}}} \\
w_2 & = & 1 - w_1 \, ,
\label{eq:2KNweights}
\end{eqnarray}
where the second equation simply shows that the weightings are true fractions summing to unity.

We see from eqs.~(\ref{eq:2KNe}) and (\ref{eq:2KNweights}) that the observed two-KN signature will always lie between the results of the individual events, as one might expect.  Thus, if there are two KNe at different times, say 10 and 20 Mya, then the result would interpolate between the first two points seen in the KN panels of Fig.~\ref{fig:rpro-regolith}.  We expect that our results span the likely range of possibilities, so realistic multiple-KN cases will lie between the lowest and highest points.  This is also the case when there are more than two KNe.

\section{The Cosmic-Ray Background of Radioisotopes in Lunar Regolith}
\label{sec:background}

Cosmic rays with MeV energies and above undergo reactions in the lunar regolith and produce radioisotopes.  This cosmogenic component creates an irreducible background for radioisotopes produces in astrophysical explosions.  Here we determine the level of this background and outline a strategy to use regolith measurements to disentangle the cosmogenic background from the SN or KN signal we seek.  We also estimate the level of {\em  destruction} of SN and KN radioisotopes by cosmic-ray interactions.

\subsection{Cosmic-Ray Production of Radioisotopes in Lunar Regolith}
\label{subsec:production}

Cosmic rays incident on the lunar regolith
create showers of secondary  protons and neutrons.  These cascades are the most intense at depths around $\sim (10,30) \ \rm cm$ below the surface for protons and neutrons respectively and are attenuated below \cite{Reedy1972,Michel1991,Leya2001}.
For the 10-100 MeV energies most important for our purposes, the neutron fluxes are larger by more than an order of magnitude because neutrons do not suffer the Coulomb losses that stop the protons.  These secondary particles react with regolith material to produce radioisotopes, whose abundance is a background that competes with our SN or KN signal.

For radioisotope $i$ produced by cosmic-ray interactions of target nucleus $j$, 
the production rate per unit volume
is 
\begin{equation}
q_i = \avg{\Phi_p \sigma_{pj \rightarrow i}+\Phi_n \sigma_{nj\rightarrow i}} n_j    
= \Gamma_{j\rightarrow i}^{\rm CR} n_j
= \Gamma_{j\rightarrow i}^{\rm CR} \frac{\rho}{A_j m_{\rm u}} X_j \, .
\end{equation}
This is the product of the target number density $n_j$
and the cosmic-ray interaction the rate per target $\avg{\Phi \sigma}$,
with $\Phi$ the cosmic-ray flux and $\sigma$ the cross section for producing species $i$,
summed over cosmic-ray protons and neutron and averaged over the cosmic-ray spectrum {\em in situ}.
We see that the cosmic-ray production 
is proportional to the target abundance, i.e,, the mass fraction:
$q_i \propto n_j \propto \rho_j \propto X_j$.

The production competes with the radioactive decay at a rate per volume, or activity $a_i = n_i/\tau_i$.
The production and decay rates are driven to an equilibrium
where $a_i^{\rm eq} = q_i$.  This state is reached over a timescale of order $\tau_i$, i.e.,  within a few Myr in our case, and so should be a good approximation for our purposes.  Thus we expect the cosmogenic species to have equilibrium abundances given by $n_i^{\rm eq} = q_i \tau_i$.

Because cosmic-ray production is proportional
to the target abundance, cosmogenic radioisotope measurements are often reported as a decay rate per target mass, i.e.,
the specific activity
\begin{equation}
    {\cal A}_{i|j} = \frac{a_i}{\rho_j}
    = \frac{n_i}{\tau_i X_j \rho}
    \stackrel{\rm eq}{=}
    \frac{\Phi \sigma}{A_j m_u} \, ,
\end{equation}
where $\tau_i$ is the mean life of $i$,
and $X_j = \rho_j/\rho$ is the mass fraction of the target. 
This ratio depends only on the cosmic-ray flux in the in the regolith (intensity and spectrum) and the cross sections for
radioisotope production.
We can thus find the number of cosmogenic radioisotope atoms per unit mass as
\begin{eqnarray}
\label{eq:cosmogen}
  & \frac{n_i^{\rm CR}}{\rho} & = {\cal A}_{i|j} \; X_j \tau_i \\
    \nonumber
    & = & 5.3 \times 10^5 {\rm atoms/g}  
    \pfrac{X_j}{100 \ \rm ppm}
    \pfrac{\tau_i}{10 \ \rm Myr}
    \pfrac{{\cal A}_{i|j}}{1 \ {\rm dpm/kg}j}
\end{eqnarray}
where ${\rm dpm/kg}j = {\rm decay \ min^{-1}  \ (kg} \ j)^{-1}$.  Note that a mass fraction of 1 ppm corresponds to $X(1 \, {\rm ppm}) = 10^{-6}$, so our fiducial value in eq.~(\ref{eq:cosmogen}) is for $X(100 \, {\rm ppm}) = 10^{-4}$.

We can use eq.~(\ref{eq:cosmogen})
to infer the cosmic-ray backgrounds 
for our radioisotopes of interest and
compare them with the levels we predict for the signals from nearby explosions.
For example, the cosmic-ray background for \i129 is mainly from cosmic-ray interactions with barium and tellurium isotopes. The tellurium abundance is generally small ($X_{\rm Te} \le 10^{-6}$) in the lunar regolith, while the barium abundances in the lunar soil span $X_{\rm Ba} \sim 42 - 850 \ \rm ppm$ \footnote{\url{https://www.lpi.usra.edu/lunar/samples/}}. Adopting ${\cal A}_{129\rm{I}|\rm  Ba} = 1.5 \ \rm dpm/kg \, Ba$ \citep{Schnabel2004}, we estimate the cosmogenic \i129 atoms per unit mass to be $\sim 2.4 \times10^7 \ \rm atoms/g$, shown as the \i129 background in Fig.~\ref{fig:rpro-regolith}, along with values from measurements of \i129 in a lunar rock \citep{Nishiizumi1983} and regolith \citep{Nishiizumi1989}; for these, we use a barium mass fraction $X_{\rm Ba} = 389 \ \rm ppm$ \citep{SChnetzler1971}. The background is proportional to the barium abundance and thus samples with smaller $X_{\rm Ba}$ would be ideal.

The main target isotope for the cosmogenic production of \hf182 
is the tungsten isotope \iso{W}{186}. The TALYS~\citep{TALYS2}~\footnote{\url{https://tendl.web.psi.ch/tendl_2021/talys.html}} theoretical cross-section values for cosmic rays interacting with this isotope are 
$\sigma(p+\iso{W}{186}) \simeq 0.9 \ \rm mb$, and 
$\sigma(n+\iso{W}{186}) \simeq 2 \ \rm mb$.  
We combine these with a tungsten regolith mass fraction of $X_{\rm W} = 10^{-6}$ \citep{Kruijer2017}, isotope fraction $\iso{W}{186}/W=0.28$, and a cosmic-ray flux in the regolith of $\Phi_N = 4 \ \rm cm^{-2} s^{-1}$.
This gives the \hf182 background shown as the horizontal line in Fig.~\ref{fig:rpro-regolith}.

In general, the lunar soil
will contain both an astrophysical signal from a nearby explosion (SN or KN) and a cosmogenic background.
For radioisotope $i$, the number density in a sample
$n_i^{\rm obs} = n_i^{\star} + n_i^{\rm CR}$
sums the two components,
and so we have
\begin{eqnarray}
\pfrac{e_i}{\rho}_{\rm obs}
& = & \pfrac{n_i^\star}{\rho}
+ \pfrac{n_i^{\rm CR}}{\rho} \\
\label{eq:abvsXmix}
& = & \pfrac{{\cal F}_i^\star}{\rho h_{\rm mix}} + \frac{\Gamma_{j\rightarrow i}^{\rm CR} \tau_i}{A_j m_{\rm u}} X_j \\
& = & \pfrac{{\cal F}_i^\star}{\rho h_{\rm mix}} + {\cal A}_{i|j} \tau_i X_j  \ , 
\end{eqnarray}
where the last equation assumes that the fluence ${\cal F}_i^\star$
of the isotope from a nearby explosion is mixed over a depth $h_{\rm mix}$.

\subsection{Strategy to distinguish the astrophysical signal from background in lunar regolith sample}
\label{subsec:distinguish}

Equation (\ref{eq:abvsXmix}) shows that a potential signal from an astrophysical explosion will be mixed with a cosmogenic background, and we thus require  a means of separating these components.
The available samples of lunar regolith and lunar soil show variations in elemental
abundances.  This suggests a simple strategy:  Search for species $i$  of interest
in samples with low target abundances $X_j$, and then plot $n_i/\rho$ vs $X_j$.
One can then find both the slope and intercept, which encode the cosmic-ray and astrophysical components, respectively.
This simple approach faces the problem that regolith samples at different depths experience different mixing \citep{Costello2018} and different cosmic-ray exposure \citep{Reedy1972,Michel1991,Leya2001}.  This can introduce large spreads in both the slope and intercept of our linear trend of $n_i/\rho$ vs $X_j$.  We have verified with lunar and meteoritic data that the $n(\fe60)/\rho$ versus the target $X_{\rm Ni}$ abundance that the observed trend is too noisy for a robust inference of the astrophysical component.  The uncertainties in the linear slope are large and consistent with zero.

\begin{figure}[htb]
    \centering
    \includegraphics[width=0.8\columnwidth]{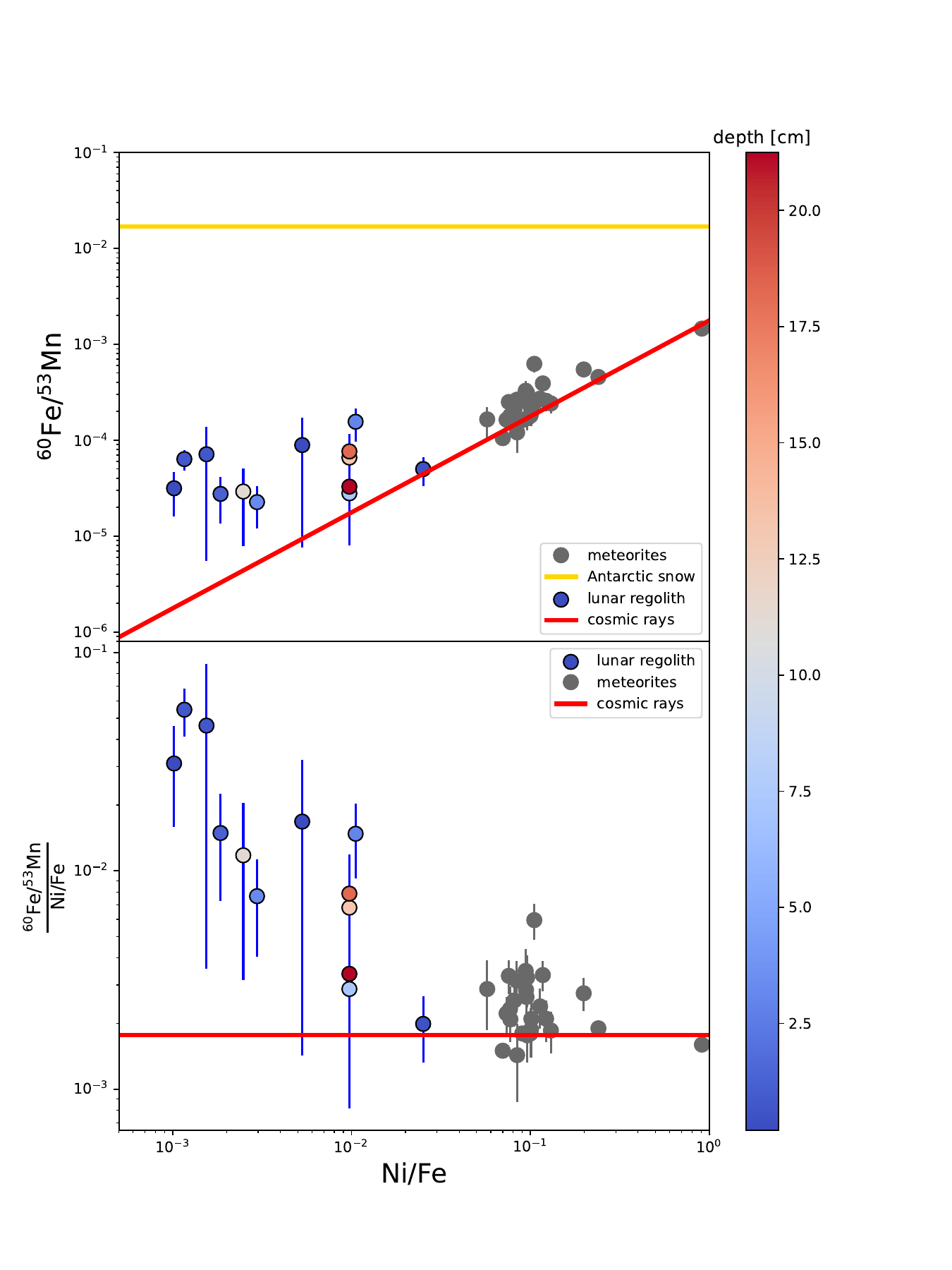}
    \caption{Ratio of $\fe60/\mn53$ for meteorites and lunar regolith.
    Panel (a): ratio of $\fe60/\mn53$ versus ${\rm Ni/Fe}$.  The linear trend for cosmic-ray production shown as a line, 
    and the excess in regolith indicates another source. 
    Panel (b): ratio of $(\fe60/\mn53)/({\rm Ni/Fe})$ versus ${\rm Ni/Fe}$: The cosmic-ray trend is horizontal, and the excess above it indicates the importance of an additional source.}
    \label{fig:60Fe/53Mn}
\end{figure}

To overcome the problem of different cosmic-ray exposures for different samples, one can measure another radioisotope in the same samples--a species that is only of cosmogenic origin. \citet{Fimiani2016}) successfully adopted this approach, measuring not only \fe60 in lunar regolith, but also \mn53, which has a high cosmogenic abundance due to the large abundance of its target nuclei such as \iso{Fe}{56}.  Here we use their lunar and meteoritic data to illustrate this procedure.  Our analysis elaborates the procedure discussed recently ~in ref.~\cite{Koll2022}.

We consider two radioisotopes: one denoted by $i$ that has or may have an SN component, e.g., \fe60, \pu244, \i129, etc.  The other is dominated by  cosmic-ray production; we denote this by $k$ and have in mind \mn53.  
Thus, both species obey eq.~(\ref{eq:abvsXmix}), but one has no discernible astrophysical component:
We use this cosmogenic species $k$ as the tracer.  
We write the number of radioisotope atoms per sample mass as $y_i = n_i/\rho$, so that
$y_i = y_i^{\rm CR} + y_i^\star$,
while $y_k = y_k^{\rm CR}$.
For the cosmic-ray components 
$y_i^{\rm CR} = \Phi_i \sigma_{j\rightarrow i} y_j \tau_i
= \tau_i {\cal A}_{i|j} X_j/A_j m_{\rm u}$, with a similar expression for species $k$.  
Using the \fe60 and \mn53 system as an example, we have
\begin{eqnarray}
\frac{\fe60}{\mn53}
& = & \frac{y_i}{y_k}
= \frac{n_i}{n_k} \\
& = & \frac{\sigma_{j\rightarrow i} \tau_i}{\sigma_{\ell \rightarrow k} \tau_j}
\frac{y_j}{y_\ell} + \frac{y_i^\star}{y_k^{\rm CR}} \\
& = & \frac{\sigma_{\rm Ni \rightarrow 60}\tau_{60}}{\sigma_{\rm Fe \rightarrow 53} \tau_{53}} 
\pfrac{\rm Ni}{\rm Fe}
+ \frac{1}{\Phi \sigma_{\rm Fe \rightarrow 53} \tau_{53}}
\frac{\fe60^\star}{\rm Fe} \, .
\end{eqnarray}
We see that a plot of 
$\fe60/\mn53$ versus Ni/Fe
should be linear with zero intercept
if there is no SN component,
with a slope that only depends on the
ratio of the $\sigma \tau$ values for the two species, not the cosmic-ray flux. 

Figure \ref{fig:60Fe/53Mn} plots $\fe60/\mn53$ versus Ni/Fe for both lunar regolith samples that should contain SN \fe60, and for meteorites that should not \citep{Fimiani2016,Koll2022}.   
We see that the meteoritic data falls on a line with a nonzero slope, indicating the presence of correlated cosmogenic production of both \fe60 and \mn53.  
The presence of a positive intercept points to an SN component, at a level that depends on the $\fe60^\star/{\rm Fe_{\rm regolith}}$ mixing of SN material in the regolith, as well as the cosmic-ray exposure $\Phi \sigma_{53} \tau_{53}$.

We can adapt this strategy to analyze the $r$-process radioisotopes with future lunar measurements.
Along with these species of interest, it is important to measure at least one cosmogenic-dominated radioisotope
such as \mn53, but it would be of interest to measure others as well such as \al26, which is abundantly produced from the spallation of silicon isotopes such as \iso{Si}{28}, as well as other examples, particularly those with lifetimes comparable to the species of interest.

Note that we have assumed that the cosmic-ray tracer species, \mn53 in our example, does not have an appreciable SN contribution in the regolith.  In fact, SN production of \mn53 should occur, and \citet{Korschinek2020} report evidence for it in a Fe-Mn crust.  But in this crust, \mn53 is still dominated by the cosmogenic component, which comes from interplanetary dust infall on Earth.  We find that in a lunar sample the cosmogenic \mn53 signal should dominate even more because the \mn53 production in the regolith column will far exceed the accreted dust fluence \citep{Love1993}.  
Thus, \mn53 can effectively be treated as purely of cosmic-ray origin.  

\subsection{Cosmic-Ray Destruction of Radioisotopes in Lunar Regolith}
\label{subsec:destruction}

Cosmic rays will also interact with the radioisotopes deposited by nearby explosions, converting them to other nuclides. This process reduces the signal we seek and so is important to estimate.  Cosmic-ray destruction of radioisotopes occurs mainly through spallation reactions.

We have examined destruction rates of \fe60, \i129, and \pu244 by cosmic-ray proton and neutron spallation on the lunar regolith. 
In general, the spallation (or inelastic) cross-section values for these isotopes from TALYS \citep[][see footnote 17]{TALYS2} are $\sigma_{\rm spall} \lesssim O(10^{4})$~mb at maximum, and in the regolith
the cosmic-ray proton flux from 10 to 100 MeV is $\Phi_p \lesssim O(10^{-1})  \ \rm cm^{-2} \, s^{-1}$ while the neutron flux at the same energies is $\Phi_n \lesssim \ O(10^{0})  \ \rm cm^{-2} \, s^{-1}$ \citep{Michel1991}.  
The destruction rate is therefore at most is about $\Gamma_{\rm inel} = (\Phi_p+\Phi_n) \sigma_{\rm inel} \lesssim O(10^{-22}) \ \rm s^{-1}$. This is far less than the radioactive decay rate of these isotopes $1/\tau_i > 10^{-16} \ \rm s^{-1}$ (take \pu244 for example, $\Gamma_{244+n} \tau_{244} \sim 4 \times 10^{-8} \ll 1$). 
Thus, radioactive decay overwhelmingly dominates the losses of these species, as we have assumed throughout. 
Consequently, the cosmic-ray destruction rates can be ignored for the lunar measurements of these isotopes.

\bibliography{SNr}
\bibliographystyle{aasjournal}

\end{document}